\documentclass{llncs}%
\usepackage{fancyheadings}
\usepackage{amsfonts}
\usepackage{colordvi} 
\usepackage{amsmath}
\usepackage{amssymb}
\usepackage{graphicx}
\ifx\pdftexversion\undefined \else \fi
\newtheorem{thm}{Theorem}[section]

\newtheorem{lem}[thm]{Lemma}

\newtheorem{prop}[thm]{Proposition}

\newcommand{\secref}[1]{Section~\ref{#1}}

\numberwithin{equation}{section}
 \numberwithin{thm}{section}
         
\newcommand\g{\gamma}

\newcommand\e{\varepsilon}
\renewcommand\l{\lambda}

\newcommand\rc{{\sc rc4}}

\newcommand\G{\Gamma}
\newcommand\f{\frac}

\newcommand{\Z}{{\mathbb{Z}}}

\renewcommand\i{^{-1}}
\renewcommand\({\left(}
\renewcommand\){\right)}

\newcommand{\bx}{\hfill$\square$\vspace{.6cm}}

\newcommand{\comment}[1]{}
\newcommand{\ignore}[1]{}
\newcommand{\rotate}{\ggg}
\newcommand{\xor}{\oplus} 
\begin{document}

\title{MV3: A new word based stream cipher using rapid mixing and revolving buffers}

\author{Nathan Keller\inst{1}\thanks{Partially supported by the Adams
Fellowship.}, Stephen D. Miller\inst{1,2}\thanks{Partially supported
by NSF grant DMS-0301172 and an Alfred P. Sloan Foundation
Fellowship.}, Ilya Mironov\inst{3}, and Ramarathnam
Venkatesan\inst{4}}

\institute{Einstein Institute of Mathematics, The Hebrew
University,  Givat Ram, Jerusalem 91904, Israel \\
\email{nkeller@math.huji.ac.il}   \vspace{.2cm} \and Department of
Mathematics, Rutgers University, Piscataway, NJ 08854 \\
\email{miller@math.rutgers.edu}
 \vspace{.2cm} \and
 Microsoft Research, Silicon Valley Campus, 1065 La Avenida,
Mountain View, CA 94043 \\ \email{mironov@microsoft.com}
 \vspace{.2cm} \and
Cryptography and Anti-piracy Group, Microsoft Research, 1 Microsoft
Way,  Redmond, WA
98052  \\ and \\
Cryptography, Security and Algorithms Research Group, Microsoft
Research India\\  Scientia - 196/36 2nd Main,
Sadashivnagar, Bangalore 560 080, India     \\
\email{venkie@microsoft.com}}

\maketitle

\begin{abstract}

\vspace{-.5cm}

\textsc{mv3} is a new \emph{word based} stream cipher for encrypting
 long streams of data. A direct adaptation of a byte based cipher
such as \rc\ into a 32- or 64-bit word version will obviously need
vast amounts of memory. This scaling issue necessitates a look for
new components and principles, as well as mathematical analysis to
justify their use. Our approach, like \rc's, is based on rapidly
mixing random walks  on directed graphs (that is, walks which reach
a random state quickly, from any starting point).  We begin with
some well understood walks, and then introduce nonlinearity in their
steps in order to improve  security and  show  long term statistical
correlations are negligible. To  minimize the short term
correlations, as well as to deter attacks using equations involving
successive outputs, we provide a method for sequencing the outputs
derived from the walk using three revolving buffers. The cipher is
fast
--- it runs at a speed of less than 5 cycles per byte on a Pentium IV processor.
A word based cipher needs to output more bits per step, which
exposes more correlations for attacks. Moreover we  seek simplicity
of construction and transparent analysis. To meet these
requirements, we use a larger state and claim security corresponding
to only a fraction of it.  Our design is for an adequately secure
word-based cipher; our very preliminary estimate puts the security
close to exhaustive search for keys of size $\leq 256$ bits.

\vspace{.1cm}

\end{abstract}



\section{Introduction}

Stream ciphers are widely used and essential in practical
cryptography. Most are custom designed, e.g.~alleged \textsc{rc4}
\cite[Ch.~16]{schneier}, \textsc{seal}~\cite{seal},
\textsc{scream}~\cite{scream}, and \textsc{lfsr}-based \textsc{nessie} %
submissions such as \textsc{lili}-128, \textsc{snow}, and
\textsc{sober}~\cite[Ch.~3]{nessie}. The \textsc{vra} cipher
\cite{vra} has many provable properties, but requires more memory
than the rest. We propose some new components and  principles for
stream cipher design, as well as their mathematical  analysis, and
present a concrete stream cipher  called {\textsc{mv3}}.

To motivate our construction, we begin  by considering \rc\ in
detail. It is an exceptionally short,  byte-based algorithm that
uses only 256 bytes of memory. It is based on random walks (card
shuffles), and has no serious attacks. Modern personal computers are
evolving from 32 to 64 bit words, while a growing number of smaller
devices have different constraints on their word and memory sizes.
Thus one may desire ciphers better suited to their architectures,
and seek designs that scale nicely across these sizes. Here we focus
on scaling up such random walk based ciphers. Clearly, a direct
adaptation of \rc\ would require vast amounts of memory.

The security properties  of most stream ciphers are not based on
some hard problem (e.g., as RSA is based on factoring). One would
expect this to be the case in the foreseeable future.
Nevertheless, they use components that -- to varying degrees --
are analyzable in some idealized sense. This analysis typically
involves simple statistical parameters such  as cycle length and
mixing time. For example, one idealizes each iteration of the main
loop  of \rc\ as a step in a random walk over its state space.
This can be modeled by a graph $G$ with nodes consisting of
$S_{256}$, the permutations on 256 objects, and edges connecting
nodes that differ by a transposition. Thus far no serious
deviations from the random walk assumptions are known. Since
storing an element of $S_{2^{32}}$ or $S_{2^{64}}$ is out of the
question, one may try simulations using smaller permutations;
however, this is nontrivial if we desire both competitive speeds
and a clear analysis. It therefore is attractive to consider other
options for the underlying graph $G$.

One of the most important parameters of \rc\  is its mixing time.
This denotes the number of  steps one needs to start from an
arbitrary state and achieve uniform distribution over the state
space through a sequence of independent random moves. This parameter
is typically not easy to determine. Moreover, \rc\ keeps a loop
counter that is incremented modulo 256, which introduces a memory
over 256 steps. Thus its steps are not even Markovian (where a move
from the current state is independent of  earlier ones).
Nevertheless, the independence of moves has been a helpful
idealization (perhaps similar to  Shannon's random permutation model
for block ciphers), which we will also adhere to.

We identify and focus on the following problems:

\renewcommand{\labelitemi}{$\bullet$}

\begin{itemize}
  \item{\bf Problem 1 -- Graph Design.} How to design graphs whose random walks are suitable for stream
  ciphers that work on arbitrary word sizes.
  \item{\bf Problem 2 -- Extraction.} How to extract
  bits to output from (the labels of)
   the nodes visited by walk.
  \item{\bf Problem 3 -- Sequencing.} How to sequence the nodes visited by the walk
  so as to diminish any attacks that use relationships (e.g. equations) between successive
  outputs.
\end{itemize}

We now expand on these issues.  At the outset, it is important to
point out the desirability of simple register operations, such as
additions, multiplications, shifts, and \textsc{xor}'s.  These are
crucial for fast implementation, and preclude us from using many
existing constructions of expander graphs (such as those in \cite
{LPS,expnotes}). Thus part of the cipher design involves new
mathematical proofs and constructions.  The presentation of the
cipher does not require these details, which may be found in
Appendix~\ref{background}.

{\bf High level Design Principles:} Clearly, a word based cipher has
to output more bits per step of the algorithm. But this exposes more
relationships on the output sequence, and to mitigate its effect we
increase the state size and aim at security that is only a fraction
of the log of the state size. We also tried to keep our analysis as
transparent and construction as  simple as possible. Our key
initialization is a bit bulky and in some applications may require
further simplifications, a topic for future research.

\subsection{Graph Design: Statistical properties and Non-linearities}

In the graph design, one wants to keep the mixing time $\tau$ small
as a way to keep the long term correlations negligible.  This is
because many important properties are guaranteed for walks that are
longer than $\tau$.  For example, such a walk visits any given set
$S$ nearly the expected number of times, with exponentially small
deviations (see Theorem~\ref{gilthm}).  A corollary of this fact is
that each output bit is unbiased.

Thus one desires the optimal mixing time, which is on the order of
$\log N$,  $N$ being the size of the underlying state space.  Graphs
with this property have been well studied, but the requirements for
stream ciphers are more complicated, and we are not aware of any
work that focuses on this issue.  For example, the  graphs whose
nodes are $\Z/2^n\Z$ (respectively $(\Z/2^n\Z)^*$) and  edges are
$(x,x+g_i)$ (respectively $(x,x\cdot g_i)$),  where $g_i$ are
randomly chosen and $i=O(n)$,  have this property \cite{ar}. While
these graphs are clearly very efficient to implement, their
commutative operations are  quite linear and hence the attacks
mentioned in Problem 3 above  can be effective.

To this end, we introduce some  nonlinearities into our graphs.
For example, in the graph
on $\Z/2^n\Z$ from the previous paragraph, we can also add edges of the form
$(x,hx)$ or $(x,x^r)$.  This intuitively allows for more randomness, as well as disrupting
relations between successive outputs.  However, one still needs to
 prove that the mixing time of such a modified graph is still
small.  Typically this type of analysis is hard to come by,
 and in fact was previously believed to be false.
However, we are able to give rigorous proofs in some cases, and
 empirically found the numerical evidence
to be stronger yet in the other cases.  More details can be found in
the Appendices.

 Mixing up the
random walks on multiplicative and additive abelian groups offers a
principled way to combine with nonlinearities for an effective
defense. As a practical matter, it is necessary to ensure that our
(asymptotic) analysis applies when parameters are small, which we have verified experimentally.

We remark here that introduction of nonlinearities was the main
motivation behind the construction of the $T$-functions of Klimov
and Shamir (\cite{ks1}). They showed that the walk generated by a
$T$-function deterministically visits every $n$-bit number once before
repeating. A random walk does not go through all the nodes in the
graph, but the probability that it returns to a previous node in $m$
steps tends to the uniform probability at a rate that drops
exponentially in $m$. It also allows us to analyze the statistical
properties as indicated above.  (See
Appendix~\ref{background} for more background.)

\subsection{Extraction}

Obviously, if the nodes are visited  truly randomly, one can simply output
the {\sc
lsb}'s of the node, and extraction is trivial.  But when there are
correlations among them, one can base an attack on studying equations
involving successive outputs.
One solution to this problem is to  simultaneously hash a number of
successive nodes using a suitable hashing function, but
 this will be expensive since the hash function has to work
  on very long inputs.

 Our solution to the sequencing
problem below allows us to instead hash a linear combination of the
nodes in a faster way.  A new aspect of our construction is that our
hash function itself evolves on a random walk principle.  We apply
suitable rotations on the node labels (to alter the internal states)
at the extraction step to ensure the top and bottom half of the
words mix well.

\subsection{Sequencing}

As we just mentioned,  the sequencing problem becomes significant if
we wish to hash more bits to the output (in comparison to \rc).
First we ensure that our graph is directed and has no short cycles.
But this by itself is insufficient, since nodes visited at steps in
an interval $[t,t+\Delta]$, where $\Delta\ll\tau$, can have strong
correlations. Also, we wish to maximize the number of terms required
in equations involved in  the attacks mentioned in Problem 3. To
this end, we store a short sequence of nodes visited by the walk in
buffers, and sequence them properly. The buffers ensure that any
relation among output bits is translated to a relation involving
many nonconsecutive bits of the internal state. Hence, such
relations cannot be used to mount efficient attacks on the internal
state of the cipher.

The study of such designs appear to be of independent interest.
 We are able to justify
their reduction of correlations via a
theorem of \cite{CHJ02-masking} (see Section~\ref{ss:lda}).

\subsection{Analysis and Performance}

We do not have a full analysis of the exact cipher that is
implemented. However, we have ensured that our idealizations are in line
with the ones that allow \rc\ be viewed via random walks.
Of course some degree of idealization is necessary because random bits are required to
implement
 any random walk; here our design resembles that of alleged \textsc{rc4} \cite[Ch.~16]{schneier}.
Likewise, our cipher involves combining steps from
different, independent random walks on the same underlying graph.
We are able to separately analyze these processes, but
although
 combining such steps should intuitively only enhance randomness,
our exact mathematical models hold only for
 these separate components and hence we performed numerical tests as
 well.

Our cipher  \textsc{mv3} is fast on 32 bit processors --- it runs at
a speed of 4.8 cycles a byte on Pentium IV, while the speed of
\textsc{rc4} is about 10 cycles a byte.  Only two of the eSTREAM
candidates \cite{Performance} are faster on similar architecture.

We evaluated it against some known attacks and we present the details in Section~\ref{s:security}.
 We note that some of the guess-and-determine attacks
against \textsc{rc4} (e.g.~\cite{backtrack}) are also applicable
against \textsc{mv3}. However, the large size of the internal state
of \textsc{mv3} makes these attacks much slower than exhaustive key
search, even for very long keys.

The security claim of \textsc{mv3} is that no attack faster than
exhaustive key search can be mounted for keys of length up to 256
bits.\footnote{Note that \textsc{mv3} supports various key sizes of
up to 8192 bits. However, the security claims are only for keys of
size up to 256 bits.}

The paper is organized as follows: In Section~\ref{s:design} we give
a description of \textsc{mv3}. Section~\ref{s:rationale} contains
the design rationale of the cipher. In Section~\ref{s:security} we
examine the security of \textsc{mv3} with respect to various methods
of cryptanalysis. Finally, Section~\ref{s:summary} summarizes the
paper.  We have also included appendices giving some mathematical
and historical background.

\section{The Cipher MV3}
\label{s:design}

In this section we describe the cipher algorithm and its basic
ingredients. The letters in its name stand for ``multi-vector'', and
the number refers to the three revolving buffers that the cipher is
based upon.

{\bf Internal state.} The main components of the internal state of
\textsc{mv3} are three revolving buffers $A$, $B$, and $C$ of length
32 double words (unsigned 32-bit integers) each and a table $T$ that
consists of 256 double words. Additionally, there are publicly known
indices $i$ and $u$ ($i\in[0\dots 31]$, $u\in[0\dots 255]$), and
secret indices $j$, $c$, and $x$ ($c,x$ are double words, $j$ is an
unsigned byte).

Every 32 steps the buffers shift to the left: $A\leftarrow B$,
$B\leftarrow C$, and $C$ is emptied. In code, only the pointers get
reassigned (hence the name ``revolving'', since the buffers are
circularly rotated).

{\bf Updates.} The internal state of the cipher gets constantly
updated by means of pseudo-random walks. Table $T$ gets refreshed
one entry every 32 steps, via application of the following two
operations:
\begin{tabbing}
\hspace{5cm}\=$u\leftarrow u+1$\+\\
$T[u] \leftarrow T[u] + (T[j] \rotate 13)$.
\end{tabbing}
(Symbol $x \rotate a$ means a circular rotation to the right of the
double word $x$ by $a$ bits).

In other words, the $u$-th element of the table, where $u$ sweeps
through the table in a round-robin fashion, gets updated using
$T[j]$.

In its turn, index $j$ walks (in every step, which can be idealized
as a random walk) as follows:
\begin{tabbing}
\hspace{5cm}\=$j \leftarrow j \,   \, + (B[i]  \text{~mod~}256),$
\end{tabbing}
where $i$ is the index of the loop. Index $j$ is also used to update
$x$:
\begin{tabbing}
\hspace{5cm}\=$x \leftarrow  x+T[j],$
\end{tabbing}
which is used to fill buffer $C$ by $C[i] \leftarrow (x \rotate 8)$.

Also, every 32 steps the multiplier $c$ is additively and
multiplicatively refreshed as follows:
\begin{tabbing}
\hspace{5cm} \=$c \leftarrow  c+(A[0]\rotate 16)$\+\\
\=$c \leftarrow c \vee 1$\+\\
\=$c\leftarrow c^2\ \ \text{(can be replaced by} \ \, c\leftarrow c^3)$\\
\end{tabbing}

{\bf Main loop.} The last ingredient of the cipher (except for the
key setup) is the instruction for producing the output. This
instruction takes the following form:
\begin{tabbing}
\hspace{2.5cm}output:\hspace{1.1cm}  $(x \cdot c) \, \xor \,
A[9i+5] \, \xor \, (B[7i+18] \rotate 16)$.
\end{tabbing}
The product $x \cdot c$ of two 32-bit numbers is taken modulo
$2^{32}$.

Putting it all together, the main loop of the cipher is the
following:
\begin{tabbing}
\hspace{2.5cm}\=\textbf{Input:} \quad\=length \textit{len}\+\\
\textbf{Output:} \>stream of length \textit{len}\\
\hspace{1cm}\=\textbf{repeat }$\textit{len}/32$ \textbf{times}\+\\
\hspace{1cm}\=\textbf{for }\=$i=0$ \textbf{ to } 31\+\+\\
    $j \leftarrow  j + (B[i] \bmod 256)$\\
    $x \leftarrow  x+T[j]$\\
    $C[i] \leftarrow  (x \rotate 8)$\\
    output $(x \cdot c)\xor A[9i+5] \xor (B[7i+18] \rotate 16)$ \-\\
    \textbf{end for}\\
    $u\leftarrow u+1$\\
    $T[u]\leftarrow  T[u] + (T[j] \rotate 13)$ \\
    $c\leftarrow c+(A[0]\rotate 16)$\\
     $c \leftarrow c \vee 1$\\
    $c\leftarrow c^2\ \ \text{(can be replaced by} \ \, c\leftarrow c^3)$\\
    $A\leftarrow B$, $B\leftarrow C$\-\\
\textbf{end repeat}
\end{tabbing}

{\bf Key initialization.}

The key initialization algorithm accepts as inputs a key $K$ of
length \textit{keylength}, which can be any multiple of 32 less than
or equal to  8192 (we recommend at least 96 bits), and an initial
vector $IV$ of the same length as the key. The key remains the same
throughout the entire encryption session, though the initial vector
changes occasionally.  The initial vector is publicly known, but
should not be easily predictable. For example, it is possible to
start with a ``random'' $IV$ using a (possibly insecure)
pseudo-random number generator known to the attacker, and then
increment the $IV$ by 1 every time (see Section~\ref{ss:tmdtoa}).

The key initialization algorithm is the following:

\begin{tabbing}
\hspace{2.5cm}\=\textbf{Input:} \quad\=key \textit{key} and initial
vector
\textit{IV}, both of length \textit{keylength} double words\+\\
\textbf{Output:} internal state that depends on the key and the $IV$\\
\hspace{1cm}\=$j,x,u \leftarrow  0$\+\\
$c \leftarrow  1$\\
fill $ A,B,C,T $ with $0x\mathtt{EF}$\\
\textbf{for }\=$i=0$ \textbf{ to } 3\+\\
    \textbf{for }\=$l=0$ \textbf{ to } 255\\
        \>$T[i+l] \leftarrow  T[i+l] + (\textit{key}[l\bmod \textit{keylength}]\rotate 8i) +
        l$.\\
    \textbf{end for}\\
    \textrm{produce 1024 bytes of \textsc{mv3} output}\\
    \textrm{encrypt $T$ with the resulting key stream}\-\\
\textbf{end for}\\
\textbf{for }\=$i=4$ \textbf{ to } 7\+\\
    \textbf{for }\=$l=0$ \textbf{ to } 255\\
        \>$T[i+l] \leftarrow  T[i+l] + (\textit{IV}[l\bmod \textit{keylength}]\rotate 8i) +
        l$.\\
    \textbf{end for}\\
    \textrm{produce 1024 bytes of \textsc{mv3} output}\\
    \textrm{encrypt $T$ with the resulting key stream}\-\\
\textbf{end for}
\end{tabbing}

Note that when only the $IV$ is changed, only the second half of the
key initialization is performed.

\section{Design Rationale}
\label{s:rationale}

In this section we describe more of the motivating principles behind
the new cipher.

{\bf Internal state.} The internal state of the cipher has a huge
size of more than 11,000 bits. This makes guess-and-determine
attacks on it (like the attack against \textsc{rc4} in
\cite{backtrack}) much slower than exhaustive key search, even for
very long keys. In addition, it also secures the cipher from
time/memory tradeoff attacks trying to invert the function $f: State
\longrightarrow Output$, even for large key sizes. More detail on
the security of the cipher with respect to these attacks appears in
Section~\ref{s:security}.

The buffers $A$, $B$, $C$ and table $T$, as well as the indices $j$,
$c$, and $x$ should never be exposed. Since the key stream is
available to the attacker and  depends on this secret information,
the cipher strictly adheres to the following design principles:
\begin{tabbing}
\hspace{1cm}\textbf{Principle 1.} Output words must depend on as
many secret words as possible.\\
\hspace{1cm}\textbf{Principle 2.} Retire information faster than
the adversary can exploit it.
\end{tabbing}

As the main vehicle towards these goals, we use random walks (or,
more precisely, pseudo-random walks, as the cipher is fully
deterministic).

{\bf Updates.} The updates of the internal state are based on
several simultaneously applied random walks. On the one hand, these
updates are very simple and can be efficiently implemented. On the
other hand, as shown in Appendix~\ref{background}, the update
mechanism allows one to  mathematically prove some randomness
properties of the sequence of internal states. Note that the random
walks are interleaved, and the randomness of each one of them relies
on the randomness of the others. Note also that the updates use
addition in $\Z/2^n\Z$ and not a bitwise \textsc{xor} operation.
This partially resolves the problem of high-probability short
correlations in random walks: In an undirected random walk, there is
a high probability that after a short number of steps the state
returns to a previous state, while in a directed random walk this
phenomenon does not exist. For example, if we would use an update
rule $x \leftarrow x \xor T[j]$, then with probability $2^{-8}$
(rather than the trivial $2^{-32}$) $x$ would return to the same
value after two steps. The usage of addition, which  unlike
\textsc{xor} is not an involution, prevents this property. However,
in the security proof for the idealized model we use the undirected
case, since the known proofs of rapid mixing (like the theorem of
Alon and Roichman \cite{ar}) refer to that case.

{\bf Introducing nonlinearity.} In order to introduce some
nonlinearity we use a multiplier $c$ that affects the cipher output
in a multiplicative way. The value of $c$  is updated using an
expander graph which involves both  addition and multiplication, as
explained in Appendix~\ref{background}. It is far from clear the
squaring or cubing operation still leaves the mixing time small and
our theorem addresses this.

Our update of $c$ involves a step $c \leftarrow c \vee 1$. This
operation may  at a first seem odd, since it leaks  {\sc lsb}$(c)$
to attacker, who may use it for a distinguishing attack based only
on the  {\sc lsb} of outputs, ignoring $c$ entirely. However, this
operation is essential, since otherwise the attacker can exploit
cases where $c=0$, which occur with a relatively high probability of
$2^{-16}$ due to the $c\leftarrow c^2$ operation (and last for 32
steps at a time).  In this situation, they can disregard  the term
$x \cdot c$ and devise a guess-and-determine attack with a much
lower time complexity than the currently possible one.

{\bf Sequencing rule.} The goals of this step were explained in
section 1.3. Our output rule is based on the following general
structure: The underlying walk $x_0,x_1,\dots,x_n,\dots$ is
transformed into the output $y_0,y_{1},\dots,y_n,\dots$ via a linear
transformation:
$$
y_i \ \ = \ \ x_{n_{i1}}\xor x_{n_{i2}}\xor\cdots \xor x_{n_{ik}}.
$$
Without loss of generality, we  assume that the indices are sorted
$n_{i1}<n_{i2}<\dots<n_{ik}$. Let $\mathcal{N}=\{n_{ij}\}$. The set
$\mathcal{N}$ is chosen to optimize the following parameters:
\begin{enumerate}
\item Minimize the latency and the buffer size required to compute
$y_i$. To this end, we require that there will be two constants $m$
and $C$, between $64$ and $256$, such that $i-C\leq n_{ij}\leq i$
for each $i\geq m$ and $1\leq j\leq k$. We additionally constrain
$n_{ik}=i$ for all $i>m$;

\item Maximize the minimal size of a set of pairs $x_i,x_{i+1}$
that can be expressed as a linear combination of $y$'s. More
precisely, we seek to maximize $a$ such that the following holds for
some $j_1,\dots,j_b>m$  and $i_1,\dots,i_a$:
\begin{equation}\label{eq:tuple}
(x_{i_1}\xor x_{i_1+1})\xor (x_{i_2}\xor x_{i_2+1})\xor\cdots\xor
(x_{i_a}\xor x_{i_a+1}) \ \ = \ \ y_{j_1}\xor y_{j_2}\xor\cdots\xor
y_{j_b}.
\end{equation}
Notice that the value of $b$ has not been constrained, since usually
this value is not too high and the attacker can obtain the required
data.
\end{enumerate}

Intuitively speaking, the second constraint ensures that if the
smallest feasible $a$ is large enough, no linear properties of the
$x$ walk propagate to the $y$ walk. Indeed, any linear function on
the $y$ walk can be expressed as a function on the $x$ walk. Since
the $x$ walk is memoryless, any linear function on a subset of $x$'s
can be written as a \textsc{xor} of linear functions on the
intervals of the walk. Each such interval can in turn be broken down
as a sum of pairs. If $a$ is large enough, no linear function can be
a good distinguisher. Note that we concentrate on the relation
between consecutive values of the state $x$, since in a directed
random walk such pairs of states seem to be the most correlated
ones.

Constructing the set $\mathcal{N}$ can be greatly simplified if
$\mathcal{N}$ has periodic structure. Experiments demonstrate that
for sequences with period 32 and $k=3$, $a$ can be as large as $12$.
Moreover, the best sequences have a highly regular structure, such
as $n_{i1}=i-(5k\bmod 16)$ and $n_{i2}=i-16-(3k\bmod 16)$, where
$k=i\bmod 16$. For larger periods $a$ cannot be computed directly;
an analytical approach is desirable.

As soon as the set of indices is fixed, $y_i$ for $i>m$ can be
output once $x_i$ becomes available. The size of the buffer should
be at least $i-n_{ij}$ for any $i>m$ and $j$. If $\mathcal{N}$ is
periodic, retiring older elements can be trivially implemented by
keeping several buffers and rotating between them. We note that
somewhat similar buffers where used recently in the design of the
stream cipher \textsc{Py} \cite{biham}.

More precisely, if we choose the period $P=32$ and $k=3$, i.e.~every
output element is an \textsc{xor} of three elements of the walk, the
output rule can be implemented by keeping three $P$-word buffers,
$A$, $B$, and $C$.  Their content is shifted to the left every $P$
cycles: $A$ is discarded, $B$ moves to $A$, and $C$ moves to $B$.
The last operation can be efficiently implemented by rotating
pointers to the three buffers.

The exact constants chosen for $n_{ij}$ in the output rule
are chosen to maximize the girth and other useful properties
of the graph of dependencies between internal
variables and the output, which is available to the attacker.

{\bf Rotations.}

Another operation used both in the output rule and in the update of
the internal state is bit rotation. The motivation behind this is as
follows: all the operations used in \textsc{mv3} except for the
rotation (that is, bitwise \textsc{xor}, modular addition and
multiplication) have the property that in order to know the $k$
least significant bits of the output of the operation, it is
sufficient to know the $k$ least significant bits of the input. An
attacker can use this property to devise an attack based on
examining only the $k$ least significant bits of the output words,
and disregard all the other bits. This would  dramatically reduce
the time complexity of guess-and-determine attacks. For example, if
no rotations were used in the cipher, then a variant of the standard
guess-and-determine attack presented in Section~\ref{s:security}
would apply.  This variant examines only the least significant byte
of every word, and reduces the time complexity of the attack to the
fourth root of the original time complexity.

One possible way to overcome this problem is to use additional
operations that do not have this problematic property, like
multiplication in some other modular group. However, such operations
slow the cipher significantly. The rotations used in \textsc{mv3}
can be efficiently implemented  and prevent the attacker from
tracing only the several least significant bits of the words.
 We note that similar techniques were used in
the stream cipher \textsc{Sosemanuk}~\cite{Sosemanuk} and in other
ciphers as well.

{\bf Key setup.} Since the bulk of the internal state is the table
$T$, we concentrate on intermingling $T$ and the pair $(key,IV)$.
Once $T$ is fully dependent on the $key$ and the $IV$, the revolving
buffers and other internal variables will necessarily follow suit.

We have specified that the  $IV$  be as long as the $key$ in order
to prevent time/memory tradeoff attacks that try to invert the
function $g: (key,IV) \longrightarrow Output$. The $IV$ is known to
the attacker but should not be easily predictable. One should avoid
initializing the $IV$ to zero at the beginning of every encryption
session (as is frequently done in other applications), since this
reduces the effective size of the $IV$ and allows for better
time/memory tradeoff attacks. A more comprehensive study of the
security of \textsc{mv3} with respect to time/memory tradeoff
attacks is presented in Section~\ref{s:security}.

We note that the key initialization phase is relatively slow.
However, since the cipher is intended for encrypting long streams of
data, the fast speed of the output stream generation compensates for it. We note that since the
$IV$ initialization phase is also quite slow, the $IV$ should not be
re-initialized too frequently.

\section{Security}\label{s:security}

\textsc{mv3} is designed to be a fast and very secure cipher. We are
not aware of any attacks on \textsc{mv3} faster than exhaustive key
search even for huge key sizes of more than 1000 bits (except for
the related key attacks in \secref{relatedkeysubsec}), but have only
made  security claims up to a 256-bit key size. In this section we
analyze the security of \textsc{mv3} against various kinds of
cryptanalytic attacks.

\subsection{Tests}

We ran the cipher through several tests. First, we used two
well-known batteries of general tests. One is Marsaglia's
time-tested  DIEHARD collection~\cite{diehard}, and the other is the
NIST set of tests used to assess AES candidates~\cite{NIST} (with
corrections as per~\cite{correction}). Both test suites were easily
cleared by \textsc{mv3}.

In light of attacks on the first few output bytes of
\textsc{rc4}~\cite{shamir,mironov}, the most popular stream cipher, we tested
the distribution of the initial double words of \textsc{mv3} (by
choosing a random 160-bit key and generating the first double word
of the output). No anomalies were found.

\textsc{rc4}'s key stream is also known to have correlations between the
least significant bits of bytes one step away from each
other~\cite{golic}. Neither of the two collections of tests
specifically targets bits in similar positions of the output's
double words. To compensate for that, we ran both DIEHARD and NIST's
tests on the most and the least significant bits of 32-bit words of
the key stream. Again, none of the tests raised a flag.

\subsection{Time/Memory/Data Tradeoff Attacks}\label{ss:tmdtoa}

There are two main types of TMDTO (time/memory/data tradeoff)
attacks on stream ciphers.

The first type consists of attacks that try to invert the function
$f:State \longrightarrow Output$ (see, for example,
\cite{biryukov}). In order to prevent attacks of this type, the size
of the internal state should be at least twice larger than the key
length. In \textsc{mv3}, the size of the internal state is more than
11,000 bits, and hence there are no TMDTO attacks of this type
faster than exhaustive key search for keys of less than 5,500 bits
length.  Our table sizes are larger than what one may expect to be
necessary to make adequate security claims, but we have chosen our
designs so that we can keep our analysis of the components
transparent, and computational overhead per word of output minimal.
We intend to return to this in a future paper and  propose an
algorithm where the memory is premium, based on different principles
for light weight applications.

The second type consists of attacks that try to invert the function
$g:(Key,IV) \longrightarrow Output$ (see, for example,
\cite{Sarkar}).  The $IV$ should be at least as long as the key --
as we have mandated in our key initialization -- in order to prevent
such attacks faster than exhaustive key search. We note again that
if the $IV$'s are used in some predictable way (for example,
initialized to zero at the beginning of the encryption session and
then incremented sequentially), then the effective size of the $IV$
is much smaller, and this may enable a faster TMDTO attack. However,
in order to overcome this problem the $IV$ does not have to be
``very random''. The only thing needed is that the attacker will not
be able to know which $IV$ will be used in every encryption session.
This can be achieved by initializing the $IV$ in the beginning of
the session using some (possibly insecure) publicly known
pseudo-random number generator and then incrementing it
sequentially.

\subsection{Guess-and-Determine Attacks}
\label{ss:attack}

A guess-and-determine attack against \textsc{rc4} appears in
\cite{backtrack}. The attack, adapted to \textsc{mv3}, has the
following form:

\begin{enumerate}
\item The attacker guesses the values of all the 32 words in buffers
$A$ and $B$ in some loop of \textsc{mv3}, and the values of $j,c,x$
in the beginning of the loop.

\item Using the guessed values of the words in $B$, the attacker traces the
value of $j$ during the whole loop.

\item Using the output stream and the guessed values, the attacker traces
the value of $x$ during the whole loop.

\item Using the update rule of $x$ and the knowledge of $j$, the attacker
gets the values of 32 words in the $T$ array. If the attacker
encounters a word whose value is already known to her, she checks
whether the values match, and if not, discards the initial guess.

\item The attacker moves on to the next loop. Note that due to the knowledge
of buffer $A$ and some of the words $T[j]$, the attacker can trace
the update of $c$ and of the $T$ register.

\item Each ``collision'' in the  $T$ array supplies the attacker with a 32-bit
filtering condition. Since the attacker started by guessing 66
32-bit words, finding 70 collisions should be sufficient to discard
all the wrong guesses and find the right one. In 10 loops we expect
to find more than 70 such collisions, and hence $2^{14}$ bits of key
stream will be sufficient for the attacker to find the internal
state of the cipher.

\item Once the attacker knows the internal state, she can compute the entire
output stream without knowing the key.
\end{enumerate}

However, the time complexity of this attack is quite large -- more
than $2^{2000}$, since the attacker starts with guessing more than
2000 bits of the state. Hence, this attack is slower than exhaustive
key search for keys of less then 2000 bits length.

\subsection{Guess-and-Determine Attacks Using the Several Least Significant
Bits of the Words}

Most of the operations in \textsc{mv3} allow the attacker to focus
the attack on the $k$ least significant bits, thus dramatically
reducing the number of bits guessed in the beginning of the attack.
We consider two reasonable attacks along these lines.

The first attack concentrates on the least significant bit of the
output words. In this case, since the least significant bit of $c$
is fixed to $1$, the attacker can disregard $c$ at all. However, in
this case the attacker cannot trace the values of $j$, and guessing
them all the time will require a too high time complexity. Hence, it
seems that this attack is not applicable to \textsc{mv3}.

The second attack concentrates on the eight least significant bits
of every output word. If there were no rotations in the update and
output rules, the attacker would indeed be able to use her guess to
trace the values of $j$ and the eight least significant bits in all
the words of the internal state. This would result in an attack with
time complexity of about $2^{600}$. However, the rotations cause
several difficulties for such an attack:

\begin{enumerate}

\item Due to the rotations, the values the attacker knows after her initial
guess are bits $24-31$ of the words in buffer $C$, bits $16-23$ of
the words in buffer $B$, and bits $0-7$ of the words in buffer $A$
(these are the bits that affect the eight \textsc{lsb}'s of the
output words). Yet the attacker still does not know the eight
\textsc{lsb}'s of the words in buffer $B$ and hence cannot find the
value of $j$.

\item If the attacker rolls the arrays to the previous loop,
 she can find the eight \textsc{lsb}'s of the words in buffer $B$. However, the attacker cannot use her guess to
get information from the previous loop. In that loop she knows bits
$0-7$ of the words in buffer $B$  and bits $16-23$ of  the words in
buffer $C$, but in order to compare the information with the eight
\textsc{lsb}'s of the output stream, she needs bits $16-23$ of the
words in buffer $B$ and bits $24-31$ of the words in buffer $C$.
Therefore, the guesses in consecutive loops cannot be combined
together.

\end{enumerate}

Hence, it seems that both of the attacks cannot be applied, unless
the attacker guesses the full values of all the words in two
buffers, which leads to the attack described in
subsection~\ref{ss:attack} (with a time complexity of more than
$2^{2000}$).

\subsection{Linear Distinguishing Attacks}
\label{ss:lda}

Linear distinguishing attacks aim at distinguishing the cipher
output from random streams, using linear approximations of the
non-linear function used in the cipher -- in our case, the random
walk.

 In~\cite{CHJ02-masking}, Coppersmith et al. developed a
general framework to evaluate the security of several types of
stream ciphers with respect to these attacks. It appears that the
structure of \textsc{mv3} falls into this framework, to which
\cite[Theorem 6]{CHJ02-masking}
directly applies:

\begin{theorem} Let $\epsilon$ be the bias of the best linear approximation one
can find for pairs $x_i,x_{i+1}$, and let $A_N(a)$ be the number of
equations of type~(\ref{eq:tuple}) that hold for the sequence
$y_m,y_{m+1},\dots$.  Then
the statistical distance between the cipher and the random string is
bounded from above by
\begin{equation}\label{eq:distbound}
\sqrt{\sum_{a=1}^N A_N(a)\epsilon^{2a}}.
\end{equation}
\end{theorem}

Note that for $\epsilon\ll 1/2$, the bound (\ref{eq:distbound}) is
dominated by the term with the smallest $a$, which equals to 12 in
our case. Since the relation between $x_i$ and $x_{i+1}$ is based on
a random walk, $\epsilon$ is expected to be very small. Since the
statistical distance is of order $\epsilon^{24}$, we expect that the
cipher cannot be distinguished from a random string using a linear
attack, even if the attacker uses a very long output stream for the
analysis.

\subsection{Related-Key Attacks and Key Schedule Considerations}
\label{relatedkeysubsec}

Related key attacks study the relation between the key streams
derived from two unknown, but related, secret keys. These attacks
can be classified into distinguishing attacks, that merely try to
distinguish between the key stream and a random stream, and key
recovery attacks, that try to find the actual values of the secret
keys.

One of the main difficulties in designing the key schedule of a
stream cipher with a very large state is the vulnerability to
related-key distinguishing attacks. Indeed, if the key schedule is
not very complicated and time consuming, an attacker may be able to
find a relation between two keys that propagates to a very small
difference in the generated states. Such small differences can be
easily detected by observing the first few words of the output
stream.

It appears that this difficulty applies to the current key schedule
of \textsc{mv3}. For long keys, an attacker can mount a simple
related-key distinguishing attack on the cipher. Assume that
$keylength = 8192/t$. Then in any step of the key initialization
phase, every word of the key affects exactly $t$ words in the $T$
array, after which the main loop of the cipher is run eight times
and the output stream is \textsc{xor}ed (bit-wise) to the content of
the $T$ array. The same is repeated with the $IV$ replacing the key
in the $IV$ initialization phase.

The attacker considers encryption under the same key with two $IV$s
that differ only in one word. Since the key is the same in the two
encryptions, the entire key initialization phase is also the same.
After the first step of the $IV$ initialization, the intermediate
values differ in exactly $t$ words in the $T$ array. Then, the main
loop is run eight times.  Using the random walk assumption, we
estimate that, with  probability $(1-t/256)^{256}$,   each of the
corresponding words in the respective $T$ arrays used in these eight
loops are equal, making
 the
output stream  equal in both  encryptions.
 Hence, with
probability $(1-t/256)^{256}$, after the first step of the $IV$
initialization the arrays $A$, $B$, and $C$ are equal in both
encryptions and the respective  $T$ arrays differ only in $t$ words.

The same situation occurs in the following three steps of the $IV$
initialization. Therefore, with probability
\begin{equation}
(1-t/256)^{256} \cdot (1-2t/256)^{256} \cdot (1-3t/256)^{256} \cdot
(1-4t/256)^{256}
\end{equation}
all of the corresponding words used during  the entire
initialization phase are equal in the two encryptions. Then with
probability $(1-4t/256)^{32}$ all of the corresponding words used in
the first loop of the key stream generation are also equal in the
two encryptions, resulting in two equal key streams. Surely this can
be easily recognized by the attacker after observing the key stream
generated in the first loop.

In order to distinguish between \textsc{mv3} and a random cipher,
the attacker has to observe about
\begin{equation}
M = (1-t/256)^{-256} \cdot (1-2t/256)^{-256} \cdot (1-3t/256)^{-256}
\cdot (1-4t/256)^{-256} \cdot (1-4t/256)^{-32}
\end{equation}
pairs of related $IV$s, and for each pair she has to check whether
there is equality in the first $32$ key stream words. Hence, the
data and time complexities of the attack are about $2^{10} M$. For
keys of length at least $384$ bits, this attack is faster than
exhaustive key search. Note that (somewhat counter intuitively) the
attack becomes more efficient as the length of the key is increased.
The attack is most efficient for $8192$-bit keys, where the data
complexity is about $2^{10}$ bits of key stream encrypted under the
same key and $2^{15}$ pairs of related $IV$s, and the time
complexity is less than $2^{32}$ cycles. For keys of length at most
$256$ bits, the data and time complexities of the attack are at
least $2^{618}$ and hence the related-key attack is much slower than
exhaustive key search.

If we try to speed up the key schedule by reducing the number of
loops performed at each step of the key schedule, the complexity of
the related-key attack is reduced considerably. For example, if the
number of loops is reduced to four (instead of eight), the
complexity of the related-key attack becomes
\begin{equation}
M' = (1-t/256)^{-128} \cdot (1-2t/256)^{-128} \cdot
(1-3t/256)^{-128} \cdot (1-4t/256)^{-128} \cdot (1-4t/256)^{-32}
\end{equation}
In this case, the attack is faster than exhaustive key search for
keys of length at least $320$ bits. If the number of loops is
further reduced to two, the complexity of the attack becomes
\begin{equation}
M'' = (1-t/256)^{-64} \cdot (1-2t/256)^{-64} \cdot (1-3t/256)^{-64}
\cdot (1-4t/256)^{-64} \cdot (1-4t/256)^{-32}
\end{equation}
and then the attack is faster than exhaustive search for keys of
length at least $224$ bits.

If the key schedule is speed up by inserting the output of the eight
loops into the $T$ array, instead of \textsc{xor}ing it bit-wise to
the content of the $T$ array (as was proposed in a previous variant
of the cipher), the complexity of the related-key attack drops to
\begin{equation}
M''' = ((1-t/256)^{-256})^4
\end{equation}
In this case, the attack is faster than exhaustive key search even
for $256$-bit keys.

Hence, the related-key attack described above is a serious obstacle
 to speeding up the key schedule. However, we note that
the related-key model in general, and in particular its requirement
of obtaining a huge number of encryptions under different
related-$IV$ pairs, is quite unrealistic.

\subsection{Other Kinds of Attacks}

We subjected the cipher to other kinds of attacks, including
algebraic attacks and attacks exploiting classes of weak keys.
  We did not find any discrepancies in these cases.
\textBlack

\section{Summary}
\label{s:summary}

We have proposed a new fast and secure stream cipher, \textsc{mv3}.
The main attributes of the cipher are efficiency in software, high
security, and its basis upon clearly analyzable components.

The cipher makes use of  new rapidly mixing random walks, to ensure
the randomness in the long run. The randomness in the short run is
achieved by revolving buffers that are easily implemented in
software, and break short correlations between the words of the
internal state.

The cipher is word-based, and hence is most efficient on 32-bit
processors. On a Pentium IV, the cipher runs with a speed of 4.8
clocks a byte.

\subsection*{Acknowledgements}

We thank Adi Shamir for his  generous  discussions. We are grateful
to Nir Avni and Uzi Vishne for their  careful reading and  comments
on an earlier version, and Rebecca Landy for providing numerics on
expander graphs.

\appendix

\section{Appendix: Mathematical Background}
\label{background}

The good long term randomness properties of the internal state of
\textsc{mv3} are achieved by updates using
rapidly mixing random walks. Actually, the walks are only
pseudo-random since the cipher is fully deterministic, but we desire
the update rule to be as close as possible to a random walk.
 In this
appendix we recall some mathematics used to study  random walks,
such as expander graphs
and the rapid mixing property.
Afterwards, we describe two particular types of
random walks used in the \textsc{mv3} cipher: a well-known random walk in the
additive group $\Z/2^n\Z$, and a novel random walk that mixes
addition with multiplication operations.

\subsection{Rapidly Mixing Random Walks and Expander Graphs}

Recall that a random walk on a graph starts at a node $z_0$, and at
each step moves to a node connected by one of its adjacent edges at
random. A lazy random walk is the same, except that it stays at the
same node with probability $1/2$, and otherwise moves to an adjacent
node at random. Intuitively, a random walk is called ``rapidly
mixing'' if, after a relatively short time, the distribution of the
state of the walk is close to the uniform distribution ---
regardless of the initial distribution of the walk.

Next, we come to the notion of expander graph. Let
$\G$ be an undirected $k$-regular graph on $N<\infty$ vertices. Its
adjacency operator acts on $L^2(\G)$ by summing the values of a
function at the neighbors of a given vertex:
\begin{equation}\label{adjc}
    (Af)(x) \ \ = \ \ \sum_{x\sim y}f(y)\,.
\end{equation}
The spectrum of $A$ is contained in the interval $[-k,k]$.  The {\em
trivial eigenvalue} $\l=k$ is achieved by the constant eigenvector;
if the graph is connected then this eigenvalue has multiplicity 1,
and all other eigenvalues are strictly less than $k$.  A sequence of
$k$-regular graphs (where the number of vertices tends to infinity)
is customarily called a sequence of {\em expanders} if all
nontrivial eigenvalues $\l$ of all the graphs in the sequence
satisfy the bound $|\l|\le k-c$ for an absolute constant $c$.  We
shall take a slightly more liberal tack here and consider graphs
which satisfy the weaker eigenvalue bound $|\l|\le k-c(\log N)^{-A}$
for some constant $A\ge 0$.

The importance of allowing the  lenient  eigenvalue bound $|\l|\le
k-c(\log N)^{-A}$ is that a random walk on such a graph mixes in
$\operatorname{polylog}(N)$ time, even if $A>0$. More precisely, we
have the following estimate (see, for example, \cite[Proposition
3.1]{jmv}).

\begin{prop}\label{expander-prop}
\label{rapmix} Let $\G$ be a regular graph of degree $k$ on $N$
vertices. Suppose that the eigenvalue $\lambda$ of any nonconstant
eigenvector satisfies the bound $|\lambda|\le \sigma$ for some
$\sigma<k$. Let $S$ be any subset of the vertices of $\G$, and $x$ be
any vertex in $\G$. Then a random walk of any length at least $
\frac{\log{2N/|S|^{1/2}}}{\log{k/\sigma}}$ starting from $x$ will
land in $S$ with probability at least $\frac{|S|}{2N} =
\frac{|S|}{2|\G|}$.
\end{prop}

\noindent Indeed, with $\sigma=k-c(\log N)^{-A}$, the random walk
becomes evenly distributed in the above sense after $O((\log
N)^{A+1})$ steps.

Next, we come to the issue of estimating the
 probability that the random walk returns to
a previously visited node.  This is very important for cryptographic
purposes, since short
cycles lead to relations which an attacker can exploit.
The following result gives a very precise estimate of how unlikely
it is that a random walk returns to the vertex it starts from.
More
generally, it shows that if one has {\em any} set $S$ consisting of,
say, one quarter of all nodes, then the number of visits of the
random walk to this set will be exceptionally close to that of a
purely random walk in the sense that it will obey a Chernoff type
bound. This in turn  allows one to show that the idealized cipher passes
all the moment tests.

\begin{thm}\label{gilthm}
(\cite[Theorem~2.1]{gillman}) Consider a random walk on a
$k$-regular graph $\G$ on $N$ vertices for which the second-largest
eigenvalue of the adjacency operator $A$ equals $k-\e k$, $\e>0$.
Let $S$ be a subset of the vertices of $\G$, and $t_n$ the random
variable of how many  times a particular walk of $n$ steps along the
graph lands in $S$.  Then, as sampled over all random walks, one has
the following estimate for any $x>0$:
\begin{equation}\label{gilpunch}
    \operatorname{Prob}\left[ \  \left|t_n
    - n\f{|S|}{|\G|} \, \right|  \  \ge \ x \,\right]
      \ \ \le \ \ \(1+\f{x\e}{10n}\)
      e^{-x^2 \e /(20n)}\, .
\end{equation}
\end{thm}

Thus even with a moderately small value of $\e$, the random walk
avoids dwelling in any one place overly long.  The strength of the
Chernoff type bound (\ref{gilpunch}) is also useful for ruling out
other substitutes for random walks because of their non-random
behavior. For example, it has been shown by Klimov and Shamir
\cite{ks} that iterates of their $T$-functions on $n$-bit numbers
cycle through all $n$-bit numbers exactly once, whereas our random
walks will have  very large expected return times.

In practice, algorithms  often actually consider random walks  on {\em directed}
 graphs.
 The connection between rapid mixing of directed graphs
(with corresponding adjacency/transition matrix  $M$) and undirected graphs is
as follows.  A result of J. Fill shows that if the additive
reversalization (whose adjacency matrix is $M+M^t$)  or
multiplicative reversalization (whose adjacency matrix is $MM^t$)
rapidly mixes, then the lazy random walk on the directed version also
rapidly mixes.
 From this it is easy to derive the effect of having
    no self-loops as well.
 Moreover, if the undirected graph has
expansion, then so does the directed graph --- provided it has an
Eulerian orientation.  It is important to note that this implication can also
 be used
to greatly improve poorly mixing graphs.
For example,
we will present a graph
in Theorem~\ref{expthm} which involves
additive reversalization in an extreme case: where the original graph is
definitely not an expander (the random walk mixes only in time
proportional to the number of vertices $N$), yet the random walk on
the additive reversalization mixes in $\operatorname{polylog}(N)$
time.

 Expander graphs are natural sources of
(pseudo)randomness, and  have numerous applications as
extractors, de-randomizers, etc. (see \cite{expnotes}). However,
there are a few practical problems that have to be resolved before
expanders can be used in cryptographic applications.
One of these, as mentioned above,
 is a
serious security weakness: the walks in such a graph have a
constant probability of returning to an earlier node in constant
number of steps.
 It is possible to solve this problem by adding the current state
(as a binary string) to that of another process which has good
short term properties, but this increases the cache size.
 In addition, if the
graph has  large directed girth (i.e.~no short cycles), then the
short term return probabilities can be minimized or even eliminated.

\subsection{Additive Random Walks on $\Z/2^n\Z$}

Most of the random walks used in the cipher, namely the random
walks used in the updates of $j$, $x$, and $T$, are performed in the
additive group $\Z/2^n\Z$. The mixing properties of these walks can
be studied using results on  {\em Cayley graphs} of this
group.
 In general, given a group $G$ with a set of generators $S$,
  the Cayley graph $X(G,S)$ of   $G$ with respect to  $S$ is the graph whose vertices
  consist of elements of $G$, and whose edges connect pairs $(g,gs_i)$, for all
   $g\in G$
  and $s_i\in S$.

Alon and Roichman~\cite{ar} gave a detailed study of the expansion properties of
abelian Cayley graphs, viewed as undirected graphs.  They showed that $X(G,S)$ is an expander
when $S$ is a randomly chosen subset of $G$ whose  size is proportional
to $\log|G|$. More precisely, they have shown
the following:

\begin{thm}(\cite{ar}).
For every $0<\delta<1$ there is a positive constant $c=c(\delta)$ such that the
following assertion holds. Let $G$ be a finite abelian group, and let $S$ be a
random set of $c \log |G|$ elements of $G$. Then the expected
value of the second largest eigenvalue of the normalized adjacency
matrix of $X(G,S)$ is at most $1-\delta$.
\end{thm}

The normalized adjacency matrix is simply the adjacency matrix, divided by the
degree of the graph.  Thus, in light of the results of the last section, the
proposition implies that these random abelian Cayley graphs are  expanders,
and hence random walks on them mix rapidly.

 Using second-moment
methods it can be shown that the graph is ergodic (and also that the
length of the shortest cycle is within a constant factor of
$\log|\Gamma|)$ with overwhelming probability over the choice of
generators. The significance of this is that we need not perform a
lazy random walk, which would introduce undesirable short term
correlations as well as waste cycles and compromise the
cryptographic strength.

In \textsc{mv3}, the rapid mixing of the random walks updating $x,j$
and $T$ follows from the theorem of Alon and Roichman. For example,
consider the update rule of $x$:
$$
x \leftarrow  x+T[j],
$$
The update rule corresponds to a random walk on the Cayley graph
$X(G,S)$ where $G$ is the additive group $\Z/2^n\Z$ and $S$ consists
of the 256 elements of the $T$ register. Note that we have $|S|=4
\log_2(|G|)$. In order to apply the theorem of Alon and Roichman we
need that the elements of the $T$ array will be random and that the
walk will be random, that is, that $j$ will be chosen each time
randomly in $\{0,\ldots,255\}$. Hence, assuming that $j$ and $T$ are
uniformly distributed, we have a rapid mixing property for $x$.
Similarly, one can get rapid mixing property for $j$ using the
randomness of $x$.

\subsection{Non-linear Random Walks}

In order to introduce some nonlinearity to the cipher, we use a
multiplier $c$ that affects the cipher output in a multiplicative
way. The multiplier itself is updated using a nonlinear random walk
that mixes addition and multiplication operations. The idealized
model of this random walk is described in the following theorem:

\begin{thm}\label{expthm} Let $N$ and $r$ be relatively prime positive
integers greater than 1, and $\bar{r}$ an integer such that
$r\bar{r}\equiv 1\pmod N$. Let $\G$ be the 4-valent graph on
$\Z/N\Z$ in which each vertex $x$ is connected to the vertices
$r(x+1)$, $r(x-1)$, $\bar{r}x+1$, and $\bar{r}x-1$. Then there
exists a positive constant $c>0$, depending only on $r$, such that all
nontrivial eigenvalues $\l$ of the adjacency matrix of $\G$ satisfy
the bound
\begin{equation}\label{thmbd}
|\l| \ \ \le \ \ 4 \ - \ \f{c}{(\log N)^2}\,,
\end{equation}
or are of the form
$$\l \ \ = \ \ 4 \cos(2 \pi k/N) \ \ \ \
\text{for~}k\text{~satisfying~~}rk\equiv k \!\!\! \pmod N.$$ In
particular, if $N$ is a power of 2 and $(r-1,N)=2$, then $\G$ is a
bipartite graph for which all eigenvalues not equal to $\pm 4$
satisfy (\ref{thmbd}).
\end{thm}

The proof of the theorem can be found in Appendix~\ref{app:proof}.  The result
means that for a fixed $r$, $\G$ is an expander graph in the looser
sense that its eigenvalue separation is at least $c/(\log N)^2$ for
$N$ large. This is still enough to guarantee that the random walk on
the graph mixes rapidly (i.e.~in $\operatorname{polylog}(N)$ time).

We note that although we use an additive notation, the theorem holds
for any cyclic group, for example a multiplicative group in which
the multiplication by $r$ corresponds to exponentiation (this is the
non-linearity we are referring to). Also the expressions $r(x\pm
1)$, $\bar{r}x\pm 1$ may  be replaced by  $r(x\pm g)$, $\bar{r}x\pm
g$ for any integer $g$ relatively prime to $N$.  Additionally, the
expansion remains valid if a finite number of extra relations of
this form are added.

We also note that it was observed by Klawe \cite{klawe} that graphs
of the form described in the theorem cannot be expanders with a {\em
constant} eigenvalue separation, i.e.~the assertion of the theorem
is false without the logarithmic terms in the denominator. Even so,
this would not change the polynomial dependence of $\log N$ in the
mixing time, but only  improve its exponent.

The operation used in the \textsc{mv3} cipher algorithm itself is slightly
 different: it involves not only addition steps, but
also a squaring or cubing step.  Though this is not covered directly
the Theorem, it is similar in spirit.  We have run extensive
numerical tests and found that this operation can in fact greatly
enhance the eigenvalue separation, apparently giving eigenvalue
bounds of the form $|\l|\le \sigma$ for some constant  $\sigma<4$
(Klawe's theorem does not apply to this graph).  Thus the squaring
or cubing operations are not covered by the theoretical bound
(\ref{thmbd}), but empirically give stronger results anyhow.

\section{Appendix: Proof of Theorem~\ref{expthm}}
\label{app:proof}

We begin with some considerations in harmonic analysis.  We may
write the adjacency operator on $L^2(\G)=L^2(\Z/N\Z)$ as
\begin{equation}\label{adj}
    A \ \ = \ \ M P + P^t M  \ \ = \ \ (M P) \ + \ (MP)^t\, ,
\end{equation}
where
\begin{equation}\label{matM}
    (Mf)(a) \ \ = \ \ f(a+1) \ + \ f(a-1)
\end{equation}
and
\begin{equation}\label{matP}
    (Pf)(a) \ \ = \ \ f(ra) \ , \ \ \ (P^tf)(a) \ \ = \ \
    f(\bar{r}a)\,.
\end{equation}
The additive characters of $\Z/N\Z$ play an important role.  They
are indexed by integers $k\in \Z/N\Z$ as follows:
\begin{equation}\label{addchar}
    \chi = \chi_k \, : \ a  \ \mapsto \  e^{2\pi i k a/N}\,.
\end{equation}
These characters are eigenfunctions of $M$ with eigenvalue
$\l_\chi = \chi(1)+\chi(-1)$, so that $\l_{\chi_k}=2\cos(2\pi
k/N)$. Furthermore $P\chi=\chi^r$, which means $P\chi_k=\chi_{rk}$
and $P^t\chi_k=\chi_{\bar{r}k}$.

The operator $A$ is self-adjoint, so its spectrum may be analyzed
by means of the Rayleigh quotient. To prove the theorem, it
suffices to show the existence of a constant $c>0$ such that
\begin{equation}\label{rayleigh}
    \max_{v\perp {\bf 1}} \left |\f{\langle Av,v \rangle}{\langle v,v \rangle}\right| \ \ \le \ \
     4 \ - \ \f{c}{(\log N)^2}\,.
\end{equation}
Here ${\bf 1}$ denotes the constant function on the graph, which
is the trivial character $\chi_0$, and $\langle v,w
\rangle=\sum_{j=1}^N v_j\overline{w_j}$ denotes the $L^2$-inner
product of functions on $\G$.  Every vector $v\in L^2(\G)$ has an
expansion of the form $v=\sum c_\chi \cdot \chi$ in terms of the
basis of characters $\chi_k$ ; the condition that $v\perp {\bf 1}$
is simply equivalent to requiring that $c_{\chi_0}=0$.

Let us now calculate the inner products in (\ref{rayleigh}) for
$v=\sum_{\chi\neq{\bf 1}} c_\chi\cdot \chi$, using the fact that
$\langle \chi,\chi'\rangle = N$ if $\chi=\chi'$, and 0 otherwise.
First, $\langle v,v \rangle =\sum N|c_\chi|^2$. As
\begin{equation}\label{mchik}
    A\chi_k \ \ = \ \ MP\chi_k \ + \ P^tM\chi_k \ \ = \ \
    M\chi_{rk} \ + \ P^t\l_k \chi_k \ \ = \ \
    \l_{rk}\chi_{rk} \ +
    \ \l_k\chi_{\bar{r}k}\,,
\end{equation}
$\chi_k$ is an eigenfunction of $A$ with eigenvalue $2\l_k$ if
$k\equiv rk\pmod N$.  This accounts for the explicit eigenvalues
which are mentioned in the statement of the theorem.
 We have that
\begin{equation}\label{mvcalc}
    Av \ \ = \ \ \sum_{k\,=\,1}^{N-1} \,  c_k  \, \l_{rk}  \,  \chi_{rk}
     \ + \
     \sum_{k\,=\,1}^{N-1} \, c_k \,  \l_k \,  \chi_{\bar{r}k}\,,
\end{equation}
where we have set $c_k=c_{\chi_k}$ for notational convenience.
The
inner product $\langle Av,v \rangle $ satisfies
\begin{equation}\label{innprod}
\aligned
    \langle Av,v \rangle \ \ & = \ \ \sum_{k,\ell\,=\,1}^{N-1} \, c_k\,
    \overline{c_\ell} \left[\, \l_{rk} \langle \chi_{rk},\chi_\ell \rangle
    \ + \ \l_k \langle \chi_{\bar{r}k},\chi_\ell \rangle \,
    \right]\\
 & = \ \ N \sum_{k\,=\,1}^{N-1} \, c_k \,
  \overline{c_{rk}} \, \l_{rk} \ +
 N \sum_{\ell\,=\,1}^{N-1} \, \overline{c_\ell}
 \, c_{r\ell} \, \l_{r\ell} \\
 &
\le  \ \ N \sum_{k\,=\,1}^{N-1} \, |c_k| \,
  |c_{rk}| \, |\l_{rk}| \ +
 N \sum_{\ell\,=\,1}^{N-1}
 \, |c_{r\ell}|\, |c_\ell| \, |\l_{r\ell}| \, . \endaligned
\end{equation}
We are now reduced to a problem about quadratic forms.  For $1\le
k,\ell \le N- 1$, let
\begin{equation}\label{akldef}
    a_{k,\ell} \ \ = \ \ \left\{%
\begin{array}{ll}
         |\l_k| +  |\l_\ell| , &   ~~~k\equiv r\ell \pmod N \text{~~~and~~~}  \ell\equiv rk \pmod N
         \\
    |\l_k|, & ~~~k\equiv r\ell \pmod N \text{~~~and~~~}  \ell\not\equiv rk \pmod N
         \\
    |\l_\ell|, & ~~~k\not\equiv r\ell \pmod N \text{~~~and~~~}  \ell\equiv rk \pmod N
         \\
    0, & \hbox{~~~otherwise.} \\
\end{array}%
\right.
\end{equation}
We need to show the existence of a constant $c>0$ for which
\begin{equation}\label{needbd}
    \sum_{k,\ell=1}^{N-1} \, a_{k,\ell}\,y_k\,y_\ell \ \ \le \ \ \(4 \ - \ \f{c}{(\log
    N)^2}\) \, \sum_{k=1}^{N-1}\,y_k^2\,
\end{equation}
for any $N-1$ real numbers $y_1,\ldots,y_{N-1}$.  Since the
spectrum coming from the characters $\chi_k$ for which $rk\equiv
k\pmod N$ has already been accounted form, we may assume $y_k=0$
for such $k$, and modify (\ref{akldef}) so that
\begin{equation}\label{aklmoddef}
    a_{k,\ell} \ \ = \ \  a_{\ell,k} \ \ = \ \ 0  \ \ \ \
    \text{if~~}rk\equiv k \pmod N  .
\end{equation}
 For this
we use the following inequality.
\begin{lem}\label{jimbo}(Proposition 8 in \cite{jimbo})
Let $(a_{ij})$ be a symmetric $n\times n$ real matrix whose
entries are nonnegative.  Let $(\g_{ij})$ be an $n\times n$ real
matrix with positive entries for which $\g_{ij}\g_{ji}=1$. Then
\begin{equation}\label{jimpunch}
    \left| \sum_{i,j\le n} a_{ij} \,  y_i  \, y_j     \right|  \ \ \le \ \ \max_{i\le n} \( \sum_{j\le n}
    \g_{ij}\,a_{ij}\) \sum_{i\le n}|y_i|^2\,.
\end{equation}
\end{lem}
Since the proof is short, we have included it here.

{\bf Proof:}  Since $0 \le (\g^{1/2} y_i \pm \g^{-1/2} y_j)^2 = \g
y_i^2 + \g\i y_j^2 \pm 2 y_i y_j$, we may bound
\begin{equation}\label{jimpf1}
\aligned
 \left| \sum_{i,j\le n} a_{ij} \,  y_i  \, y_j    \right|  \ \ & \le \ \
\f{1}{2} \sum_{i,j\le n} 2 \, a_{ij} \,  |y_i|   \, |y_j|
\\
& \le \ \ \f{1}{2}  \sum_{i,j\le n}  a_{ij} \,  (\g_{ij}\, y_i^2
\, + \,
\g_{ji} \, y_j^2)\\
& = \ \ \sum_{i,j\le n}  a_{ij} \, \g_{ij} \, y_i^2\\
& \le \ \ \max_{i\le n} \( \sum_{j\le n}
    \g_{ij}\,a_{ij}\) \sum_{i\le n}|y_i|^2\,. \qquad\qquad\text{\bx}
\endaligned
\end{equation}

Now we specify which $\g_{ij}$ to use in bounding our sequence.
(In what follows we closely follow the technique of Jimbo-Maruoka
from a different example in  \cite{jimbo}.) Given an element
$i\in\Z/N\Z$, we let $||i||$
 denote the distance from $i$ to $N\Z$.  In other words, if $i$ is represented by a residue between $0$ and $N$,
 $||i||=\min\{i,N-i\}$.  For
$s\ge 1$ set
\begin{equation}\label{akdef}
    a_s \ \ = \ \ 1 \ - \ s \, \f{d}{(\log N)^2}\, ,
\end{equation} where $d$ is a small constant (depending on $r$)
which shall be chosen later.  Given an integer $m$ relatively
prime to $N$, we define $s_m$ to be the largest integer $s$ such
that $r^s$ divides $||2m||$. Since $||2m|| \le N/2$, $s=O(\log N)$
and $a_s>0$ provided $d$ is sufficiently small.
 We  set $\g_{k\ell}=1$ {\bf except} in the following
cases:\begin{center}\begin{tabular}{|c|c|c|} \hline
  &  $||2k||<N/(2r) $& $||2k||\ge N/(2r) $\\
\hline
 $||2\ell||<N/(2r)$ &  $\g_{k,rk}=a_{s_k}$ & $\g_{r\ell,\ell}=a_{s_\ell}\i$ \\
  &  $\g_{r\ell,\ell}=a_{s_\ell}\i $&  \\
\hline
 $||2\ell||\ge N/(2r) $&  $\g_{k,rk}=a_{s_k}$ & (no exceptions) \\
\hline
\end{tabular}\end{center}
This satisfies the requirement that $\g_{k,\ell}\,\g_{\ell,k}=1$.
 We will choose the constant $d$ to be smaller yet so
that each $\g_{k\ell}\le 1+\f{1}{2}(1-\cos{\pi/(2r)})$, as we may
do. To finish the proof we must now show the existence of a
constant $c>0$ so that
\begin{equation}\label{needtoshow}
\sum_{\ell\,=\,1}^{N-1}\,\g_{k,\ell}\,a_{k,\ell} \ \ = \ \
\g_{k,rk}\,|\l_{rk}| \ + \ \g_{k,\bar{r}k}\, |\l_k| \ \ \le \ \ 4
\ - \ \f{c}{(\log N)^2}
\end{equation}
for each $1\le k \le N-1$ which does not satisfy $rk\equiv k\pmod
N$.

Case I: Assume that $||2k||\ge N/(2r)$.  Then
\begin{equation}\label{caseiimps}
    \aligned
    k   \ \ & \in \ \ \left[\f{N}{4r}, \f{N}{2} - \f{N}{4r}\right] \ \cup \
    \left[\f{N}{2}+\f{N}{4r}, N-\f{N}{4r}\right]\\
  2 \pi k/N   \ \ & \in \ \ \left[\f{\pi}{2r}, \pi - \f{\pi}{2r}\right] \ \cup \
    \left[\pi+\f{\pi}{2r}, 2\pi-\f{\pi}{2r}\right]\\
    \endaligned
\end{equation}
and $|\l_k|=2|\cos(\f{2\pi k}{N})| \le 2 \cos(\f{\pi}{2r})$.  Now
the lefthand side of (\ref{needtoshow}) is bounded by
$(1+\f{1}{2}(1-\cos\f{\pi}{2r})(2+2
\cos(\f{\pi}{2r}))=4-4\sin(\f{\pi}{4r})^4$, which is bounded away
from 4 by an positive constant depending only on $r$.

 Case II: Now  assume
that $||2k||<N/(2r)$, and that $rk$ is not congruent to $k$ modulo
$N$. Using the trivial bound that $|\l_k|, |\l_{rk}| \le 2$, the
lefthand side of (\ref{needtoshow}) is bounded by
$2(\g_{k,rk}+\g_{k,\bar{r}k})$. Both cases in the first column of
the table have $\g_{k,rk}=a_{s_k}$.  If $||2\bar{r}k||\ge N/(2r)$,
then $\g_{k,\bar{r}k}=1$ and $1+a_{s_k} \le 2-d/(\log N)^2$, so
that the bound in (\ref{needtoshow}) is satisfied so long as
$c<2d$, which it may be chosen to be.

The only remaining situation is when both $||2k||< N/(2r)$ and
$||2\bar{r}k||<N/(2r)$, where the left hand side of
(\ref{needtoshow}) is bounded by $2(a_{s_k}+a_{s_{\bar{r}k}}\i)$.
Let $-N/(2r)<m<N/(2r)$ be the integer congruent to $2\bar{r}k$
modulo $N$, i.e. so that $||2\bar{r}k||=|m|$.  Then $rm\equiv
2k\pmod N$.  Yet since $-N/2<rm<N/2$, $||2k||=|rm|$.  We may
assume that $m\neq 0$, for otherwise $2k\equiv 2rk\equiv 0\pmod
N$; this implies $k\equiv rk\pmod N$ if $N$ is odd, and $k\equiv
rk \equiv N/2$ if $N$ is even (since then $r$ is odd).
 Therefore $r$ divides $||2k||=|rm|$ to
exactly one more power than it divides $||2\bar{r}k||=|m|$. Thus
$s_{\bar{r}k}=s_k-1$. Now
$$
\gathered a_{s}\i \ \ = \ \ \(1-\f{s\,d}{(\log N)^2}\)\i \ \ = \ \
1+\f{s\,d}{(\log N)^2} + O\(\(\f{s\,d}{(\log N)^2}\)^2\) \\ = \ \
1+ \f{s\,d}{(\log N)^2} + O\(\f{d^2}{(\log N)^2}\)
\endgathered
$$ since $s=O(\log
N)$. Therefore $(a_{s_k}+a_{s_{\bar{r}k}}\i)$ equals
$$
\gathered 1 \ - \ s_{k}\f{d}{(\log N)^2}  \ + \
\(1-s_{\bar{r}k}\f{d}{(\log N)^2}\)\i \ \  \qquad\qquad\qquad \\
\qquad\qquad \le  \ \ 2 \ - \ (s_{k}-s_{\bar{r}k}) \f{d}{(\log
N)^2} \ + \ O\(\f{d^2}{(\log N)^2}\),
\endgathered
$$which
is smaller than $2-c/(\log N)^2$ for some sufficiently small
$c>0$.  This concludes the proof of Theorem~\ref{expthm}.

\section{Appendix: Related Work}

\label{s:related} Theoretically, the requirements for stream ciphers
are well understood: cryptographically secure pseudo-random number
generators  (PRNG) exist if and only if one-way functions
exist~\cite{HILL}, and such a generator would be ideal as a stream
cipher. However such known constructions would yield prohibitively
slow implementations in practice. The heart of such constructions
involves a one-way function $f$ and a hard-core bit extractor
$B(x)$. If $f$ is based on an algebraic problem such as
\textsc{dlog} or \textsc{factoring}, the resulting cipher is quite
slow even when $B(x)$ is simple and constitutes outputting some bits
of $x$. If $f$ is based on block ciphers, then
$B(x,r)=\mathit{parity}(x\wedge r)$ is often based on the
Goldreich-Levin theorem \cite{goldreichlevin}. Computing this parity
bit takes on the average $\frac{n}{2}$ cycles, where $n$ is the
machine word size. One can speed this up with some precomputations
and make it into a practical algorithm with provable properties
(e.g.~the VRA cipher \cite{vra}, which has the disadvantage of
needing to store a large array  of random bits.)

Computerized methods for random number generation go back to von
Neumann~\cite{neumann}. Many designers of PRNGs used clever
techniques to control correlations between adjacent outputs of their
algorithms, but few generators needed it as badly as LFSR-based
algorithms~\cite{LFSR}. Indeed, since the Berlekamp-Massey
algorithm~\cite{massey} efficiently determines the state of an LFSR
of length $n$ given only $2n$ bits, all LFSR-based constructions
necessarily must hide the LFSR's exact output sequence.

Historically, the first method to hedge LFSR's from the
Berlekamp-Massey attack was due to Geffe~\cite{geffe}. It combines
outputs of three synchronously clocked LFSR's to produce one stream
of output bits, using one of them as a multiplexer. This is a lossy
combiner in the sense it outputs only one of three bits generated by
the LFSR's. It was broken by Siegenthaler~\cite{correlation}, who
also broke another 3-way non-linear combiner of~Bruer~\cite{bruer}.
Another attempt by Pless~\cite{jk} ---  to make use of non-linear
J-K flip-flops to combine eight LFSR's into one key stream ---  was
broken shortly thereafter~\cite{jkbroken}.

A recommended approach to designing LFSR-based ciphers is the
shrinking generator~\cite{shrinking}. It outputs only one quarter of
its generated bits, but has proved to be secure after 10 years of
wide use and extensive scrutiny.

Most combiners considered in the  LFSR literature  are constructed
from two building blocks: a (non)linear function that mixes inputs
of several generators (this function may either be memoryless or
stateful,  though usually of very small memory), and a clocking rule
that controls the clock of some LFSR's. None of them uses deep
buffers or tries to space LFSR's outputs using schemes with
guaranteed properties. For attacks on combiners with small memory
(up to 4 bits) see~\cite{courtois03}.

A different approach for combining generators' outputs is called
{\em randomization by shuffling} \cite[Ch. 3.2.2]{knuth}. Two
algorithms popularized by Knuth are often used in modern generators:
the ``algorithm M" or MacLaren-Marsaglia algorithm~\cite{mm}, and
the ``algorithm B" or Bays-Durham algorithm~\cite{BD}. Both  are
analogous to our proposal in the sense that they store the
generator's output in a buffer and output the stored elements out of
order. The fundamental difference ---  and  source of weakness ---
of both algorithms M and B is  that they only reorder elements
without modifying them. We omit the details. For example, the
Bays-Durham algorithm operators as follows:
\smallskip

\noindent\textbf{Bays-Durham Algorithm.} \newline$Y$ is an auxiliary
variable, $T$ is the size of the buffer $V$, $m$ is the range of the
generator $\langle
X_{n}\rangle$. Initially $V$ is filled with $T$ elements $X_{0}$%
,\dots,$X_{T-1}$. Iterate the following:

\begin{enumerate}
\item set $j\leftarrow\lfloor TY/m\rfloor$.

\item set $Y\leftarrow V[j]$.

\item output $Y$.

\item set $V[j]\leftarrow$ next element of $\langle X_{n}\rangle$.
\end{enumerate}

Since the position of the output element is completely determined by
the previous element, the construction does not improve
cryptographic properties of the cipher.  If $Y$ is chosen by an
independent process (as in the algorithm M), there is still a $1/T$
chance that two elements $X_{i}$ and $X_{i+1}$ will end up next to
each other in the output sequence. More generally, the distance
between $X_{i}$ and $X_{i+1}$ is distributed according to a
geometric distribution and has average $T$. Depending on the
generator, this property may be exploitable.

Klimov and Shamir~\cite{ks} proposed a class of invertible mappings
$\{0,1\}^{n}\rightarrow\{0,1\}^{n}$ called $T$-functions that allow
introduction of non-linearity using elementary register operations
$(\vee,\wedge,\oplus,\ast,$
$+,-,x\mapsto\overline{x},x\mapsto-x,\ll)$.  The $T$-functions are
particularly well suited for fast software implementations.
An example of such a function is $f(x)=x+(x^{2}\vee5)$ $(\operatorname{mod}%
2^{n})$, for which the sequence $x_{i+1}=f(x_{i})$ spans  the entire
domain in one cycle. Each iteration requires only $3$ cycles.
Nevertheless, by choosing
$n=64$ and outputting the top half of $x_{i}$ (i.e. $H(x_{i})=$%
\textsc{msb}$_{32}(x_{i})$), they discovered that the resulting
pseudo-random sequence passed the statistical test suite for
\textsc{aes} candidates with significance level $\alpha=0.01,$ which
is better than some of the \textsc{aes} candidates. Surprisingly,
the best known cryptanalytic attacks take time $2^{cn},$ where $c$
is a constant. These attacks depend on using the structure of the
iterated output: this structure is important for proving the
properties of these functions, and slightly altering the
construction would destroy the properties. These functions allow
some of their parameters be chosen at random subject to certain
constraints.

The methods in this paper allow us to resist such attacks better,
with minimal overhead, and extend the length of the underlying key
for the stream cipher. We do not know how to extend the known
attacks in this new model.

\end{document}